\documentstyle[]{article}
\begin{document}
\title{The Fractal Universe: From the Planck to the Hubble Scale}
\author{B.G. Sidharth$^*$\\
Centre for Applicable Mathematics \& Computer Sciences\\
B.M. Birla Science Centre, Adarsh Nagar, Hyderabad - 500 063 (India)}
\date{}
\maketitle
\footnotetext{$^*$Email:birlasc@hd1.vsnl.net.in; birlard@ap.nic.in;\\
Phone:009140235081; Fax:009140237266.}
\begin{abstract}
We examine the fractal structure of the physical universe from the large
scale to the smallest scale, including the phenomenon of fractal scaling.
This is explained in terms of a stochastic underpinning for the laws of
physics. A picture in pleasing agreement with experiment and observation
at all scales emerges, very much in the spirit of Wheeler's "Law Without
Law". It is argued that our depiction of the universe is akin to a broad
brush delineation of a jagged coastline, the Compton wavelength being
comparable to the thickness of the brush strokes.
\end{abstract}
\begin{flushleft}
\end{flushleft}
\section{Introduction}
The universe is fractal and inhomogeneous on the scale of galaxies and
clusters of galaxies\cite{r1,r2}. However whether this is true at
even greater scales is debatable\cite{r3}. At the other end of space
scales, it appears that the dimension at the Planck length is less than that
of the embedding space\cite{r4}. Staying purely in the realm of
physics, it appears that the three dimensionality of space is valid at the
scale of the earth and the solar system\cite{r5}. Infact as Nicolis and
Prigogine put it\cite{r6}, "Our physical world is no longer
symbolised by the stable and periodic planetary motions that are at the
heart of classical mechanics". Even at these scales, if we step out of the
realm of physics, the usual concepts of dimensionality are no longer
valid.\\
In the words of Dyson\cite{r7} "Classical mathematics had its roots in the
regular geometric structures of Euclid and the continuing dynamic structures
of Newton. Modern mathematics began with Cantor's Set Theory and Peano's
Set-Filling Curve.... The same pathological strucutres that the mathematicians
invented to break loose from 19th Century naturalism turn out to be inherent
in familiar objects all around us".\\
We will confine ourselves strictly to the domain of physics and argue in
this condensed communication that the fractal characteristic, whether it be in the large scale
structure of the universe or in the behaviour of quarks or Planck masses are
symptomatic of an underpinning Brownian behaviour and what has been
called by Mandelbrot, scaling fractals\cite{r8}.
\section{The Fractal Universe}
As is known the distribution of galaxies is highly inhomogeneous, displaying
at large scales, structures in the form of long filaments, chains and
cellular structures: The distribution has fractal properties\cite{r9}. though, as
pointed out in the introduction it is still debatable whether this is true
at the largest scales\cite{r2,r3,r1}. Infact at large scales, the
mass distribution at distance $R$ is given by,
\begin{equation}
M (R) \sim R^\beta, 1.2 \leq \beta \leq 1.3\label{e1}
\end{equation}
(\cite{r2,r3}).\\
It was pointed out by Sidharth and Popova that\cite{r10} this is suggestive
of asymptotic two dimensionality of space and explains cosmological puzzles
like the age of the universe and age of stars anamoly and dark matter.\\
In other words the universe on a large scale resembles the Fournier or
Charlier model of the universe\cite{r8}. Physically, this had been
explained by Hoyle (cf.ref.\cite{r8}) in terms of the formation of galaxies and
stars by a cascading process in which a primordeal homogeneous gaseous cloud
becomes unstable and contracts and in the process splits into five clouds of
equal size and so on.\\
We would now like to point out that this is explained by the fact that
there is a Brownian underpinning to the
above fractal structure. As noted by Mandelbrot\cite{r8}, "the most
useful fractals involve chance and both their regularities and their
irregularities are statistical. Also the shapes described here tend to be
scaling, implying that the degree of their irregularity and / or fragmentation
is identical at all scales". He goes on to quote Nobel Laureate Jean Perrin,
to cite Brownian motion as an example of a natural fractal.\\
Indeed, as is well known in the case of Brownian motion, we have the
relation\cite{r11},
\begin{equation}
R \sim l \sqrt{N}\label{e2}
\end{equation}
where $R$ is the overall size of the system and $N$ is the number of steps
and $l$ is the length of a typical step.\\
In the context of cosmology, (\ref{e2}) is the well known Eddington relation, $R \sim 10^{28}cms$
being the radius of the universe, $N \sim 10^{80}$ being the number of
elementary particles and $l \sim 10^{-12}cms$  the
Compton wavelength of the typical elementary particle, namely a pion. This
relation has been shown to be a natural consequence of a fluctuational
cosmology by Sidharth \cite{r12,r13}. We will return
to this point later in Section 4.\\
We now observe that the following relations hold:
\begin{equation}
R \approx l_1 \sqrt{N_1}\label{e3}
\end{equation}
\begin{equation}
R \approx l_2 \sqrt{N_2}\label{e4}
\end{equation}
\begin{equation}
l_2 \approx l \sqrt{N_3}\label{e5}
\end{equation}
where in (\ref{e3}) $N_1 \sim 10^{6}$ is the number of superclusters and $l_1 \sim 10^{25}cms$
is a typical supercluster size; in (\ref{e4}) $N_2 \sim 10^{11}$ is the number
of galaxies in the universe, and $l_2 \sim 10^{23}cms$ is the typical size of
a galaxy and $N_3$ in (\ref{e5}) is the number of pions in a typical
galaxy.\\
The equations (\ref{e3}),(\ref{e4}) and (\ref{e5}) bear striking resemblance
to the equation from Brownian motion, (\ref{e2}) and tell the whole story including
fractal scaling.
They also explain the low dimensional structures of galaxies and superclusters.
Furthermore in the fluctuational cosmological scheme referred to above (cf. also
Section 4), we have,
\begin{equation}
G = \frac{a}{\sqrt{N}}\label{e6}
\end{equation}
where $G$ is the gravitational constant and $a \sim 10^{32}$.\\
Introducing (\ref{e6}) in the well known formula for the velocity $v$ at the
edges of galaxies\cite{r14} viz.,
\begin{equation}
v^2 = G \frac{M_g}{l_2},\label{e7}
\end{equation}
we get, as required, $v \sim 300$ kilometers per second: Rotational
velocities, do not tend to zero as one would expect from (\ref{e7}) for
example, but rather tend to the above constant value, thus explaining this
observational puzzle.\\
It must also be emphasized that within this framework\cite{r13}
puzzling equations like (\ref{e1}) and (\ref{e2}) also follow as a consequence
of the theory, as will be seen in Section 4.\\
Thus the underlying Brownian character in the universe is brought out,
consistently with observed data, and this explains the observed large scale
fractal structure of the universe. However it is interesting to note that an equation
like (\ref{e2}) or (\ref{e5}) does not hold for  individual stars. This is
because the gravitational force within a star is large enough to inhibit
the Brownian motion.
\section{Elementary Particles}
Starting with the Brownian or Random Walk relation (\ref{e2}), we now argue
that it is possible to deduce the Dirac equation and thence a model for all
fundamental particles namely the quarks and the leptons. A step in the
direction of such a stochastic description for the Schrodinger and Klein-
Gordon equations was taken by Nelson, Gaveau, De Pena and others \cite{r15,r16,r17}.
While Ord\cite{r18} tried a low dimensional formulation of the Dirac equation.\\
We first observe that the Compton wavelength $\hbar/mc$ which comes from
(\ref{e2}) leads to the Compton time $\hbar/mc^2$\cite{r19,r20}.
Indeed if $\Delta x$ does not tend to zero, but $\Delta t$ could, then,
velocities of infinite magnitude would be possible. So from (\ref{e2}), using
the fact that $R = cT$, where $T$ is the age of the universe, we get
\begin{equation}
T = \sqrt{N}\tau,\label{e8}
\end{equation}
where $\tau = l/c$ is the pion Compton time. Equation (\ref{e8}) is ofcourse
consistent with data and will be deduced alternatively in Section 4.
Equations (\ref{e2}) and (\ref{e8}) show that the
Compton scale is a fundamental unit of space time. Indeed it was shown that
if there is an ultimate break to the scaling, the Compton scale emerges as
this ultimate scale. This will be discussed in Section 5.  One could then easily show
that quantized space time could be considered to be more fundamental than
Planck's energy quanta\cite{r19}. Quantized space time itself has a long
history\cite{r21,r22,r23}. As T.D. Lee observes,
"space time continuum is but an approximation." Snyder\cite{r21} showed
that discrete space time is compatible with Special Relativity and
deduced equations like
\begin{equation}
[x,p_x] = \imath \hbar [1+(a/\hbar)^2 p^2_x],\label{e9}
\end{equation}
where $p^\mu$ denotes the four momentum and $a$ is the fundamental length,
the Compton wavelength in our case. We observe that as $a \to 0$, (\ref{e9})
leads to the usual quantum mechanical commutation relations.\\
We now briefly indicate how the origin of the Dirac equation lies in the
equation (\ref{e9})\cite{r24,r25}. We
consider a linear transformation of the wave function, under an infinitesimal coordinate shift in
Minkowski space. As is well known, this gives
\begin{equation}
\psi' (x_j) = [1+ l\epsilon \left(l\epsilon_{ljk}x_k\frac{\partial}
{\partial x_j}\right) + 0(\epsilon^2)]\psi (x_j)\label{e10}
\end{equation}
We next consider the commutation relations, taking $a$ to be the Compton
wavelength. We can easily verify that the choice
\begin{equation}
t =  \left(\begin{array}{l}
        1 \quad 0 \\ 0 \quad -1 \\
        \end{array}\right), \vec x =
   \ \  \left(\begin{array}{l}
         0 \quad \vec \sigma \\ \vec \sigma \quad 0 \\
       \end{array}\right)\label{e11}
\end{equation}
provides a representation for the coordinates $x$ and $t$ apart from any scalar
factors. Substitution of (\ref{e11}) in
(\ref{e10}) now leads to the Dirac equation,
\begin{equation}
(\gamma^\mu p_\mu - mc^2)\psi = 0\label{e12}
\end{equation}
Thus we obtain a rationale for
spin and the Dirac matrices in a simpler and more physical manner.
Once the Dirac equation (\ref{e12}) is deduced, it is well known that the
Schrodinger equation follows as a non relativistic approximation\cite{r26}.\\
Further, at the Compton scale, the negative energy two spinor $\chi$ of
the full four rowed Dirac spinor begins to
dominate. Moreover under reflections, $\chi$
behaves like a psuedo-spinor\cite{r26},
$$\chi \to - \chi$$
that is as a
density of weight $n = 1,$ so that\cite{r27},
\begin{equation}
\frac{\partial \chi}{\partial x^\mu} \to \frac{1}{\hbar} [\hbar \frac{\partial}
{\partial x^\mu} - n\hbar \Gamma_\sigma^{\mu \sigma}] \chi\label{e13}
\end{equation}
$\Gamma$'s being the usual Christoffel symbols.\\
We can easily identify the electromagnetic
four potential in (\ref{e13}).
The fact that $n = 1$ explains why the charge is discrete. We can also
immediately see the emergence of the metric tensor and the resulting
potential.\\
We now use the fact that the metric tensor $g_{\mu \nu}$ resulting from (\ref{e13})
satisfies an inhomogenous Poisson equation\cite{r28}, whence
\begin{equation}
g_{\mu \nu} = G \int \frac{\rho u_\mu u_\nu}{|\vec r - \vec r'|}
d^3\vec r\label{e14}
\end{equation}
where now we require the volume of integration to be the Compton volume.
As shown elsewhere\cite{r27,r13,r29,r30},
given the linearized equation of General Relativity,
(\ref{e14}) was the starting point of a geometrized formulation of Fermions
leading to the Kerr-Newman metric and which explains the remarkable and supposedly
coincidental fact that the Kerr-Newman metric describes the field of an
electron including the anomalous gyro magnetic ratio $g = 2$.\\
All this was also
shown to lead to a unified description of electromagnetism, gravitation
and strong interactions\cite{r29,r30}.\\
We now show how a unified description of quarks and
leptons can be obtained from (\ref{e14}) and how the concept of fractal
dimensioinality is tied up with it.\\
From (\ref{e13}) and (\ref{e14}) we get
\begin{equation}
A_0 = G \hbar \int \frac{\partial}{\partial t} \frac{(\rho u_\mu u_\nu )}
{|\vec r - \vec r'|} d^3r \approx \frac{ee'}{r}\label{e15}
\end{equation}
for $|\vec r - \vec r'| > >$ the Compton wavelength where $e' = e$ is the
test charge.\\
Further, from (\ref{e15}), as in the discrete case,
$d \rho u_\mu u_\nu = \Delta \rho c^2 = mc^2$ and $dt = \hbar/
mc^2$, we get
$$A_0 = \frac{e^2}{r} \sim G \frac{\hbar}{r} \frac{(mc^2)^2}{\hbar}$$
or
\begin{equation}
\frac{e^2}{Gm^2} \sim 10^{40} (\sim \sqrt{N})\label{e16}
\end{equation}
(\ref{e16}) is the well known but hitherto purely empirical relation expressing
the ratio of the gravitational and electromagnetic strengths here
deduced from theory.\\
If however in (\ref{e14}) we consider distances of the order of the Compton wavelength,
it was shown that we will get instead of (\ref{e15}), a QCD type potential
\begin{eqnarray}
4 \quad \eta^{\mu v} \int \frac{T_{\mu \nu} (t,\vec x')}{|\vec x - \vec x' |} d^3 x' +
(\mbox terms \quad independent \quad of \quad \vec x), \nonumber \\
+ 2 \quad \eta^{\mu v} \int \frac{d^2}{dt^2} T_{\mu \nu} (t,\vec x')\cdot |\vec x - \vec x' |
d^3 x' + 0 (| \vec x - \vec x' |^2) \propto - \frac{\propto}{r} + \beta r\label{e17}
\end{eqnarray}
where $T_{\mu \nu} \equiv \rho u_\mu u_\nu$. Equation (\ref{e17})
can lead to a reconciliation
of electromagnetism and strong interactions \cite{r30}. For this we need to
obtain a formulation for quarks from the above considerations. This
is what we will briefly recapitulate.\\
The doubleconnectivity or spin half of the electron
leads naturally to three dimensional space\cite{r31}, which however
breaks down at Compton scales and so we need to consider
two and one dimensions. Using the well known fact that
each of the $\rho u_\imath u_j$ in (\ref{e15}) is given by $\frac{1}{3} \epsilon$
\cite{r14}, $\epsilon$ being the energy density, it follows immediately
that the charge would be $\frac{2}{3} e$ or $\frac{1}{3}e$
in two or one dimensions, exactly as for quarks. At the same time as we are
now at the Compton scale, these fractionally charged particles
are confined as is expressed by the confining part of the QCD
potential (\ref{e17}). Further, at the Compton scale, as noted earlier
we encounter predominantly the negative energy components of the Dirac spinor
with, opposite parity. So these quarks would show neutrino type handedness,
which indeed is true.\\
Thus at one stroke, all the peculiar empirical characteristics of the quarks
for which as Salam had noted\cite{r32}, there was no theoretical rationale,
can now be deduced from theory. We can even get the correct order of
magnitude estimate for the quark masses \cite{r30}.
On the other hand neutrinos have vanishingly small mass. So their Compton wavelength is very large and by
the same argument as above, we encounter predominantly the negative energy
components of the Dirac spinor which have opposite parity, that is the
neutrinos display handedness.\\
Thus handedness and fractional charge are intimately tied up with dimensionality.
\section{Cosmological Considerations}
We will now briefly consider a fluctuational cosmological scheme which leads
back to equations (\ref{e1}) and (\ref{e2}), which were so far empirical
starting points. We consider what has been called a pre universe or quantum
vaccuum, which is a Zero Point Field type of medium, of the kind used in
considerations of stochastic electrodynamics\cite{r17}. Such
a medium was the starting point in Prigogine's cosmology\cite{r33,r34}
and in Steady State Cosmology\cite{r14}, and particles are irreversibly
created by instability or fluctuation. As Prigogine put it \cite{r34},
"The Big Bang was an event associated with an instability within the medium
that produced our universe... We consider the Big Bang an irreversible process
par excellence from a pre universe that we call quantum vaccuum."\\
Given $N$ particles in the universe, at an instant, we use the well known fact
that $\sqrt{N}$ particles are fluctuationally created. In our case the "instant"
is ofcourse the minimum time interval $\tau$, the Compton time. So we have,
$$\frac{dN}{dt} = \frac{\sqrt{N}}{\tau},$$
On integration we get (\ref{e8}) which was deduced from a different viewpoint
in Section 3.
By a similar argument we can deduce (\ref{e2}) and also \cite{r12,r13}:
\begin{equation}
\frac{Gm}{lc^2} = \frac{1}{\sqrt{N}}\label{e18}
\end{equation}
\begin{equation}
H = \frac{c}{l} \frac{1}{\sqrt{N}} \approx \frac{Gm^3c}{\hbar}\label{e19}
\end{equation}
where $H$ is the Hubble Constant and $m$ is the pion mass.\\
The model describes an over expanding universe, as infact latest observations
confirm \cite{r35}. It deduces from the theory, the mass of
the universe in terms of microphysical parameters like the mass of the pion
and Planck's Constant\cite{r36}, since, consistently $M = Nm \sim 10^{56}gm$.\\
Further equation (\ref{e18}) gives the correct value of the gravitational constant
from theory while (\ref{e19}) also deduces the Hubble constant correctly,
and moreover gives the otherwise adhoc and empirical relation between the
Hubble constant and the pion mass, which was termed mysterious by Weinberg and
others. Moreover the above scheme is
consistent with a cosmological constant $\Lambda \leq O (H^2)$ \cite{r37}
in agreement with observations. Finally the famous Large
Number relations like (\ref{e2}), all follow.\\
We next observe that the background Zero Point Field gives the correct spectral
density\cite{r17}, $\rho (\omega)\propto \omega^3$,
whence from the total intensity of radiation from the fluctuating field due to
a single star it follows that over large scales the
total mass of the universe is given approximately by\cite{r13}
$$M \propto R,$$
in agreement with equation (\ref{e1}). So the fractal low dimensionality is
a consequence.\\
We have thus been lead in this scheme of random fluctuations to our starting
point of a fractal universe as reflected by equations (\ref{e1}) and
(\ref{e2}).
\section{Discussion}
As we indicated in Section 3 there is a scaling symmetry but these scaling
fractals \cite{r8} are to be considered in a statistical sense. Infact
the dynamical origin of mass in a self similar chain was considered by
Sidharth and Altaisky\cite{r5} and it was deduced that there would be a
constant, say, $\hbar$ if scale invariance was broken at the
step $l_{\mbox{break}}$ given by
\begin{equation}
\hbar = 2^{1/2}m_ol_oc^{3^-2l_{break}}\label{e20}
\end{equation}
What is very interesting here is that identifying the constant $\hbar$ as
being proportional to 
the Planck constant, we recover from the (\ref{e20}) the Compton wavelength.
Beyond this however it would not be possible to probe further - indeed that
would lead to a contradiction in view of Heisenberg's Uncertainity Principle,
as we would have to deal with arbitrarily large energies and momenta.
This contradiction has been recognized, but Physics has lived with it\cite{r38,r13}.
All this brings us back to the Random Walk equation (\ref{e2})
and the quantized space time picture described above. Indeed these minimum
space time cut offs are very much in the spirit of Wheeler's Law Without
Law \cite{r39}. As he put it, "all of Physics in my view, will be seen someday
to follow the pattern of thermodynamics and statistical mechanics, of
regularity based on chaos, of "law without law". Specially, I believe that
everything is built higgledy-piggledy on the unpredictable outcomes of
billions upon billions of elementary quantum phenomena, and that the laws and
initial conditions of physics arise out of this chaos by the action of a
regulating principle, the discovery and proper formulation of which is the
number one task..." It may be mentioned that Wheeler's travelling salesman
problem leads to a statistical minimum length \cite{r40} which can be shown
to be the Compton wavelength itself\cite{r19}.\\
What we are doing here is, finding a thick brush in the spirit of the Richardson
effect of measuring a jagged coastline, which length would in the limit of
arbitrarily small lengths become infinite \cite{r8}. The thickness of the
brush, the Compton wavelength is Wheeler's or Mandelbrot's optimum scale
\cite{r8,r38}.\\
Finally it may be mentioned that the (\ref{e20}) leads to the fundamental
relation (\ref{e16})
It is interesting to observe that in the cosmological scheme described in
Section 4, at the epoch with $N \sim 1$, (\ref{e16}) gives the Planck mass.
At that stage all energy was gravitational as can be seen from (\ref{e16})
with the right side put equal to $1$. This would describe the Planck
mass, which indeed is a minimum Schwarschild Black Hole.\\
So the Planck particles were created in the very early epoch reminiscent of
the Prigogine cosmology. At the present epoch however we have electromagnetism
and gravitation as given by (\ref{e16}).

\end{document}